\documentstyle[12pt]{article}
         \makeatletter
         \def\thefigure{\@arabic\c@figure}\def\fps@figure{tbp}
         \def\ftype@figure{1}\def\ext@figure{lof}
         \def\fnum@figure{\protect\footnotesize Fig.\ \thefigure}
         \def\thetable{\@arabic\c@table}
         \def\fps@table{tbp}\def\ftype@table{2}\def\ext@table{lot}
         \def\fnum@table{\protect\footnotesize Table \thetable}
         \makeatother
\begin{document}
\newcommand{\tphi}{\tilde{\phi}}
\newcommand{\lton}{\stackrel{\large <}{\sim}}
\newcommand{\gton}{\stackrel{\large >}{\sim}}
\newcommand{\beq}{\begin{equation}}
\newcommand{\eeq}[1]{\label{#1} \end{equation}}
\newcommand{\beqar}{\begin{eqnarray}}
\newcommand{\eeqar}[1]{\label{#1} \end{eqnarray}}
\newcommand{\gfm}{{\rm GeV/Fm}^3}
\newcommand{\half}{{\textstyle \frac{1}{2}}}
\newcommand{\vx}{{\bf x}}
\newcommand{\vq}{{\bf q}}
\newcommand{\vp}{{\bf p}}
\newcommand{\vk}{{\bf k}}
\newcommand{\vK}{{\bf K}}
\newcommand{\vv}{{\bf v}}
\newcommand{\kn}{ $K_0$ }
\newcommand{\knb}{ $\overline{K}_0$ }
\newcommand{\ks}{ $K_s$ }

\begin{flushright}
CU-TP-670\\
IFT-P.036/94
\end{flushright}
\vspace{1cm}
 \begin{center}
{\Large  Resolving Power of 2-D Pion Interferometry$^*$
 \footnotetext{
*This work was supported by the Director, Office of Energy Research,
Division of Nuclear Physics of the Office of High Energy and
Nuclear Physics of the U.S. Department of Energy under Contract No.
DE-FG02-93ER40764.}}\\[5ex]

Sandra S. Padula$^{1}$ \footnotetext{ 1. Permanent address:
IFT/UNESP, Rua Pamplona 145, 01405-900
S\~ao Paulo, SP, Brazil.\\
%\footnotetext{ 2.
Partially supported by Conselho Nacional de Desenvolvimento
Cient\'\i fico e Tecnol\'ogico (CNPq), Brazil} and
Miklos Gyulassy
\\[2ex]
Physics Department\\
Columbia University\\
New York, N.Y. 10027\\[2ex]
\today
\end{center}
\begin{quote}  \begin{small}
Abstract:
A $\chi^2$ analysis  is performed  to test the resolving power
of two-dimensional pion interferometry using for illustration
the preliminary E802 data on $Si+Au$ at 14.6 AGeV/c.
We find that the resolving power to distinguish
two decoupling geometries of different dynamical models
is enhanced by studying the variation of the mean  $\chi^2$
per degrees of freedom
with respect to the range of the analysis in
the $q_T,q_L$ plane. The preliminary data seem to rule out
dynamical models with significant $\omega,\eta$  resonance formation yields.
 \\[2ex]
 \end{small}   \end{quote}

In this letter, the  results  of a new multi-dimensional
pion interferometric analysis of preliminary E802 $\pi^-\pi^-$
correlation data\cite{e802_pipi}-\cite{morse} on central $Si+Au$ reactions
at 14.6 AGeV/c are reported.
To study the resolving power of multi-dimensional - in contrast
to the projected one-dimensional interferometry -
we consider two dynamical scenarios that predict similar correlation functions
even though the underlying decoupling geometries differ considerably.
We use the  non-ideal inside-outside cascade model of Refs.
\cite{pgg,pg_nioc}. In one scenario long lived resonances
are neglected while in the other the  resonance
fractions as taken from the Lund model.
At higher SPS energies\cite{pgg}, we showed that long lived resonance
production predicted by the Lund Model led to
similar correlation functions to those expected
from models with quark-gluon plasma formation\cite{pratt0}, since
long lived resonances could mimic the long slow burn of a plasma.
Therefore another goal of the present study is to see if a similar ambiguity
 could exist at the lower AGS range.
There is some
fragmentary data\cite{resdat} indicating that
resonance production could be  suppressed at the AGS,
the Lund model  predicts similar resonance yields at the two
energy ranges. We show below that the
 two-dimensional $\chi^2$ interferometric analysis has  sufficient
resolving power
to rule out significant long lived resonance production in the AGS domain,
if the preliminary data
are confirmed.

Recall  that under idealized conditions
the  correlation function,
$C_2(\vk_1,\vk_2)$, of two identical
pions probes  the decoupling or freeze-out space-time
distribution of pions, $\rho(x)$, through
$C_2(\vk_1,\vk_2)= 1 + | \rho(k_1 - k_2) |^2$.
However, in actual high energy reactions, final state interactions,
correlations between coordinate and momentum variables,
and resonance production
distort this ideal interference pattern due to Bose-Einstein symmetry
(see e.g. \cite{pgg}-\cite{chapman}).
The  correlation function in this more general case can be  expressed
as\cite{pgg}
\beq
C(k_1,k_2) = \Upsilon(q)\left(
1 + \frac{|G(k_1,k_2)|^2}{G(k_1,k_1) G(k_2,k_2)} \right)
\; \; , \eeq{cce}
where $\Upsilon(q)  = (q_c/q) / ( e^{q_c/q} -1)$ is the  Gamow factor
 that distorts the pattern due to
Coulomb effects, with $q_c = 2\pi
\alpha m$ and $q=(-(k_1-k_2)^2)^{1/2}$.
The  complex amplitude, $G(k_1,k_2)$,
is  the smoothed  Fourier
transform of
the decoupling  phase-space
 distribution $
 D(x,p) = \langle \delta^4(x-x_f) \delta^4(p-p_f) \rangle
$,
where  $(x_f,p_f)$ are the  phase-space coordinates of
the pions  at their decoupling times $x^0_f$.
For minimum uncertainty
Gaussian packets,
\beq
G(k_1,k_2) =
\int d^4 x d^4 p \; D(x,p) e^{iqx}e^{q^2 \Delta x^2/2}
 e^{( p-K)^2/2\Delta p^2}
 \propto \langle e^{i q x_{f}}
e^{- K p_{f}/\Delta p^2}
\rangle
\; \; , \eeq{cegij}
where  $\Delta p $ is the rms momentum width of the packets.
 The interference pattern  depends then
not  only on the break-up  phase-space distribution, $D(x,p)$,
but in general also
on the quantal production dynamics characterized here by
$\Delta p$. The standard notation $q=k_1-k_2$ and $K=(k_1+k_2)/2$
is used throughout.
The distortions of the interference pattern due
to long lived resonances
modify the amplitude as \cite{pgg}.
\beq
G(k_1,k_2) \approx \langle  \sum_{r} f(\pi^-/r)
 \left(1-iq u_r/ \Gamma_r \right)^{-1}
\exp(i q x_r -K p_r/ \Delta p^2)\rangle
\; \; , \eeq{ceresgij}
where $f(\pi^-/r)$ is the fraction of the observed $\pi^-$'s arising from
the decay of a resonance
of type $r$, which freezes-out with final four velocity
 $u_r^\mu$.
The single inclusive pion distribution in this notation is
$P_1(\vk) \propto G(k,k)$.
The  Lund model in the AGS range suggests
that $f(\pi^-/\omega)=0.16$,
$f(\pi^-/K^*)=f(\pi^-/(\eta+\eta^\prime))=0.09$,
while some data seem to favor smaller fractions \cite{resdat}.
An aim of the present study is to test if
multidimensional pion intereferometry
can discriminate between these possibilities.

We  fit to the correlation data using
 the following
parameterization of the break-up distribution\cite{pgg,pg_nioc}
\beq
D(x,p) \propto \;\tau
 \exp\left\{-\frac{\tau^2}{ \Delta \tau^2}-\frac{(y-y^*)^2}{2 Y_c^2}
-\frac{(\eta - y)^2}{2 \Delta \eta^2}
-\frac{x_T^2}{R_T^2}\right\}  \delta(E - E_\vp)
\delta^2(\vp_T)
\; \;, \eeq{niioc}
where $\tau=(t^2-z^2)^\half$
is the freeze-out proper time, and
$\eta = \half \log((t+z)/(t-z)), y = \half\log((E+p_z)/(E-p_z))$
are the space-time and momentum rapidity variables.
The correlation between these rapidities is estimated \cite{pgg}
from the Lund model
to be  $\Delta \eta \approx 0.8$.

The goal of the present interferometry analysis is to extract
the rms transverse radius, $R_T$, at decoupling and the rms decoupling
proper time interval, $\Delta \tau$.
Note that unlike in classical cascade models\cite{rqmd,arc},
the transverse momentum in the above model is assumed to arise entirely
from  the finite momentum spread of the pion  wave-packets
through the parameter $\Delta p$.
The $y^*,Y_c,\Delta p$ are fixed by fitting the
single pion inclusive data\cite{e802_incl} with
\beq
P_1(k)\propto G(k,k)\propto e^{-\frac{(y-y*)^2}{2Y_c^2}}
e^{-m_T/T} m_T^{-1/2},
\eeq{p1}
where $T=\Delta p^2/m$ is the effective transverse mass slope.
We search for the optimal values of $R_T$ and
$\Delta \tau$ by fixing
$\Delta\eta=0.8$ from \cite{pgg,pg_nioc}
and $Y_c=0.7$, $y^*_{cm}=0$ and $\Delta p^2/m=0.17$ GeV
from the inclusive pion distributions\cite{e802_incl}.
Note that we assume  implicitly that the chaoticity parameter
$\lambda=1$  in our analysis
because, as we have emphasized in \cite{pgg,pg_nioc},
fits to data treating  $\lambda$ as a free parameter
distort the geometrical
scales in a further uncontrolled way.

For illustration, we recall that for special kinematics
 with $m_{1 T}=m_{2 T}=m_T$ and $\Delta y=
y_1-y_2$ small, the correlation function
has the approximate analytic form when resonance production is neglected
\cite{pg_nioc}
\beq
C(k_1,k_2) \approx  1 + \left\{1+\frac{1}{2} m_T^2
\Delta \tau^2 \Delta y^2 (\Delta \eta^2 + \frac{Y_c^2\Delta p^2}{\Delta p^2+
m m_T Y_c^2}) \right\}^{-2}
\; \;.\eeq{nicdtyceta}
This shows that the interference pattern
as a function of the rapidity difference depends  not
only on the geometrical quantity of interest, $\Delta \tau$,  but also on the
width, $\Delta \eta$, of the $\eta-y$ correlation as well on the
width, $Y_c$, of the finite rapidity distribution and on the
wave-packets size.

To compare theoretical correlation functions with data
projected onto two of the six dimensions, we must compute the
projected correlation function
\beq
C_{proj}(q_T,q_L) = \frac{ \int d^3 k_1 d^3 k_2
P_2(\vk_1,\vk_2) \;A_2(q_T,q_L;\vk_1,\vk_2)}
{\int d^3 k_1 d^3 k_2
P_1(\vk_1)P_1(\vk_2) \;A_2(q_T,q_L;\vk_1,\vk_2)
}
\; \; ,
\eeq{c35}
where $P_1$ and $P_2$ are the one and two pion inclusive distributions,
and $A_2$ is the experimental two pion binning and acceptance function.
It is important to note that because of this projection,
Coulomb corrections do not in general factor out of the integral.
All calculation were performed using the Monte Carlo importance sampling
method adopted in the CERES program\cite{pgg,pg_nioc}.
The acceptance function for the E802 experiment was approximated by
\beqar
A_2(q_T,q_L;\vk_1,\vk_2) &=& A_1(\vk_1) A_1(\vk_2) \Theta(20-|\phi_1-\phi_2|)
\nonumber \\  &\;& \delta(q_L-|k_{z1}-k_{z2}|)
  \delta(q_T-|\vk_{T_1}-\vk_{T_2}|)
\; \; .\eeqar{a2}
The angles are measured in degrees and the momenta in GeV/c. The single
inclusive distribution  cuts are specified by
\beq
A_1(\vk)= \Theta(14<\theta_{lab}<28)
\Theta(p_{lab}<2.2\;{\rm GeV/c})  \Theta(y_{min}>1.5) \; \; .\eeq{a2a}

To assess  the statistical significance of
the differences between the fits obtained assuming
resonance and non-resonance dynamics, we compute
the $\chi^2$ goodness of fit on a two-dimensional
grid in the $(q_T,q_L)$ plane, binned with $\delta q_T=
\delta q_L=0.01$ GeV/c. In our preliminary analysis reported in
Ref.\cite{hipags}, we  compared the square of the difference between
the theoretical and experimental correlation functions
and found an unexpected ridge of high variance along the $q_T=q_L$
diagonal.
However,  Zajc\cite{bill} pointed out that this feature could have been
an artifact of  statistical fluctuations in  ratios of random variables,
and suggested that the following  $\chi^2$ variable
\beq
\chi^2(i,j) =\frac{ [A(i,j) - {\cal N_\chi}^{-1} C_{th}(i,j)  B(i,j)]^2 }
{ \{ [\Delta A(i,j)]^2 + [ {\cal N_\chi}^{-1} C_{th}(i,j)
\Delta B(i,j)]^2 \} }
\; \; ,\eeq{chi2}
should be used instead.
In eq. (\ref{chi2}),  ${\cal N}_\chi$ is a normalization factor
which is chosen to minimize the
average $\chi^2$ and depends on the range in the
$q_T,q_L$ plane under analysis. The indices $i,j$ refer to the
 the corresponding $q_T,q_L$
bins, in each of which the experimental correlation function
is given by
\beq
C_{exp}(i,j) = {\cal N_\chi} \times \frac{A(i,j)}{B(i,j)} \; \; ; \; \;
\Delta C_{exp}(i,j) = C_{exp}(i,j) \sqrt{ \left(
\frac{\Delta A(i,j)}{A(i,j)}\right)^2 +
\left(\frac{\Delta B(i,j)}{B(i,j)}\right)^2 }
\; \; .\eeq{cexp}
The data for the numerator $A(i,j)\pm \Delta A(i,j)$ and denominator
$B(i,j)\pm \Delta B(i,j)$
were obtained from R. Morse\cite{morse} with an understanding that
the data in this form are preliminary and subject to further
 final  analysis. We note that the published one-dimensional correlation
data in \cite{e802_pipi}
are not in a form suitable  for our multi-dimesional analysis.
Unfortunately, no final multi-dimesional
correlation data are available at this time, and we use these data only for
illustration of the method and to emphasize the need to go beyond
traditional  one-dimensional projections.

        Minimization of the average $\chi^2$ is performed
by exploring the parameter space of  $R_T$ and $\Delta \tau$ and computing
the $\langle \chi^2 \rangle$, averaging over a  30x30 grid in the relative
momentum region $q_T,q_L <0.3$ GeV/c. In the vicinity of the minimum,
$R_{T_0},\Delta\tau_0$, we determine the parameters of the quadratic surface
\beq
\langle \chi^2(R_T,\Delta\tau)\rangle =\chi^2_{min} +  \alpha (R_T-R_{T_0})^2
 + \beta (\Delta\tau-\Delta\tau_0)^2 \; \; . \eeq{parab}
Recall that the average $\chi^2$ over $N$ bins is a random variable
with unit mean and rms width $\sigma=\sqrt{2/N}$ if
the $\chi(i,j)$
are normal random variables with zero mean and unit rms width.
For large $N$,
the distribution of the mean $\chi^2$ per bin
is $P(\chi^2)\propto \exp[-(\chi^2-1)^2/2\sigma^2]$.
For the $N=900$ grid under consideration, $\sigma\approx 0.047$.

 The most direct measure of the goodness of the fit in this case is
$n_\sigma=|\langle \chi^2_{min} \rangle - 1|/\sigma$,
the number of standard deviations from unity
of the average $\chi^2$ per degree of freedom.
Inserting eq. (\ref{parab}) into the asymptotic form of the
$\chi^2$ distribution for large $N$, the likelihood
for the parameter $R_T$ to have a value near the
minimum is approximately $\propto \exp[-\alpha^2(R_T-R_{T_0})^4
/2\sigma^2]$.
We therefore estimate the error on the radius to
be $\Delta R\approx \{\sqrt{2}[\Gamma(3/4)/\Gamma(1/4)] \sigma/\alpha\}^{1/2}
\approx 0.7( \sigma/\alpha)^{1/2}$.
Similarly, the estimated error on the proper time interval is
$0.7 (\sigma/\beta)^{1/2}$.
The results  of this statistical analysis
are given in Table 1.

The lego plots of the Gamow corrected data and the
corresponding theoretical correlation functions with and without resonances,
corresponding to the optimized geometrical parameters, are shown in
Figure 1 for the restricted range
$0.005<q_T,q_L<0.125$ GeV/c.
Part a) shows the 2-D data in the above range. Part (d) shows
the corresponding experimental errors, which vary  from
$\Delta C\approx 0.1$ over most of this region to
$\Delta C\approx 0.35$ toward the right
most corner at small $q_T$ and large $q_L$. In part (b) the best fit
(with $R_T=4.6$ fm and $\Delta\tau=3.4$ fm/c from Table 1
over the wider domain $q_T,q_L< 0.3$ GeV/c)
assuming no  resonance production is shown.
In part (c) the best fit with Lund resonance fractions
($R_T=3.1$ fm and $\Delta\tau=1.6$ fm/c again over the wider range) is shown.
Finally the $\chi^2(i,j)$ for the non-resonance scenario is shown in (e),
whereas that corresponding to the full resonance fit is shown in part
(f). Similar large fluctuations close to the edge of small $q_T$
and large $q_L$ is seen in both cases. However, the overall trend
suggests that the no resonance fit
is closer to the data in the region of small
 $q$ even though the global $\chi^2$ is about the same for both fits.
%\vfill

%\newpage
\begin{center}
{\bf TABLE 1: 2D-$\chi^2$ Analysis of Pion Decoupling Geometry} \\

\vskip 0.6cm

\begin{tabular}{||c||c|c||}
\hline \hline
  &  &  \\
$\chi^2(R_{T},\Delta \tau)$  & No Resonances & LUND Resonances \\
  &  &  \\
\hline
%%  &  &  \\
\multicolumn{3}{||c||}{ E802 Data Not Gamow Corrected}\\
%\hline
%$|\chi^{2}_{min}-1|/\sigma$ & 2.7 & 2.7 \\
\hline
$\langle \chi^{2}_{min} \rangle$ & 1.125 & 1.128 \\
\hline
$R_{T_0}$  & 4.1$\pm$ 1.3 & 2.5 $\pm$ 1.3 \\
\hline
$\Delta \tau_0 $ & 3.3$\pm$ 1.0 & 1.4$\pm$ 1.0 \\
\hline
$\alpha$ & 0.014 & 0.013  \\
\hline
$\beta$ & 0.021 & 0.023 \\
\hline
\multicolumn{3}{||c||}{ E802 Data  Gamow Corrected}\\
\hline
%$|\chi^{2}_{min}-1|/\sigma$ & 2.1 & 2.2 \\
%\hline
$\langle \chi^{2}_{min} \rangle$ & 1.098 & 1.104 \\
\hline
$R_{T_0}$ & 4.6$\pm$ 0.9 & 3.1$\pm$ 1.3 \\
\hline
$\Delta\tau_0$ & 3.4 $\pm$ 0.7 & 1.6$\pm$ 1.0\\
\hline
$\alpha$ & 0.027 & 0.014  \\
\hline
$\beta$ & 0.042 & 0.023 \\
\hline\hline
\end{tabular}

\end{center}

\vskip 0.7cm

The  average $\chi^2$ per degree of freedom depends, of course,
on the range of $q$ under  analysis.
We have tested this by varying the range of the analysis to restricted $q_T,
q_L$ domains, ranging from a $1\times1$ grid
corresponding to $0.005<q_T,q_L<0.015$ GeV/c,
to $2\times2$, etc. as shown in Figure 2. However,
instead of $\langle \chi^2\rangle$ we show there the number of standard
deviations from unity of the averaged $\chi^2$ per degree of freedom.
For each $n\times n$ grid, $N=n^2$ is the number of
degrees of freedom and the standard deviation is
expected to be $\sigma=\sqrt{2}/n$.
%\surd 2/n$.
The strong dependence of the number of standard deviations from unity
as a function of
the range of the analysis is brought out clearly  in Figure 2.
We see that the optimal fit including Lund resonances
is much worse than the fit without long lived resonance distortions
when the analysis is restricted to the domain $q_L,q_T\le Q_{max}\approx
100 $ MeV/c,  where the correlation function
deviates significantly  from unity.
The two models yield similar accepatable fits in terms
of $\chi^2$ for $Q_{max}>100$ MeV/c because in that large
domain
 both predict models trivially predict nearly unit correlation functions.
Also, the mean $\langle\chi^2\rangle$ begins to deviate
strongly from unity for both models when data close to the edge
of experimental acceptance, near $q_L,q_T \approx 0.3$ GeV/c,
is included.

We conclude that multi-dimensional analysis
 has high resolving power in
the domain of physical interest, and if these data are confirmed,
the present analysis would
rule out long lived resonance production models in this
energy range.
However, we note that recent  preliminary analysis in ref. \cite{new802}
seems to give  systematically lower correlation function
values in the small $q_T,q_L$ domain, in which case the resonance
or an alternate long lived source geometry would be needed.
In any case, the results  show that
 2-D $\chi^2$ analysis can be used to
improve considerably the resolving power  of interferometry
by amplifying in a quantitative way the differences
between the data and calculations.
We note that recent comparisons of interferometric
data with  cascade models, RQMD\cite{rqmd} and ARC\cite{arc},
indicate good agreement wit one-dimensional projected correlation data.
However, the global $\chi^2$ is quite poor. The study of the
dependence of that $\chi^2$ as a function of the range in those
cases should help resolve where the models fail in a more quantitative fashion.
Only in that case  can one begin to assess whether
the data are in fact consistent with standard hadronic transport models
or require novel collective dynamics.
In general, the analysis could be  sharpened
considerably by  independent measurements of
long lived  resonance abundances to reduce
that source of interferometric distortions.

\vspace{1.1cm}
Acknowledgements: We are grateful to R. Morse for his extensive help
in making his unpublished  data files and analysis
available to us. Critical discussions with S. Nagamiya
and W. Zajc  are also gratefully acknowledged.
 \\[2ex]

\vspace{4ex}

\section*{Figures Caption}
\begin{description}

\item[Fig.1] Negative pion correlations in central Si+Au
reactions\cite{e802_pipi} as a function of transverse and
longitudinal momentum difference
$q_T,q_L$. The preliminary
E802 data corrected for acceptance and Coulomb effects
are shown in part (a).
Parts (b) and (c) show
theoretical correlation functions filtered with the E802 acceptance
for cases without and with resonance production, respectively.
The transverse radius and proper time interval characterizing the pion
decoupling distribution were obtained by minimizing the $\chi^2$
over a wider region with $q_T,q_L<0.3$ GeV/c.
Part (d) shows the distribution of $\chi^2(q_T,q_L)$
for the case with resonance
production in part (c). The average $\chi^2$ parameters are listed
 in Table 1.

\vspace{4ex}

\item[Fig.2] The number of standard deviations from unity of the $\chi^2$ per
degree of freedom as a function of the range $Q_{max}$ of the analysis in the
$q_T,q_L$ plane.

\end{description}
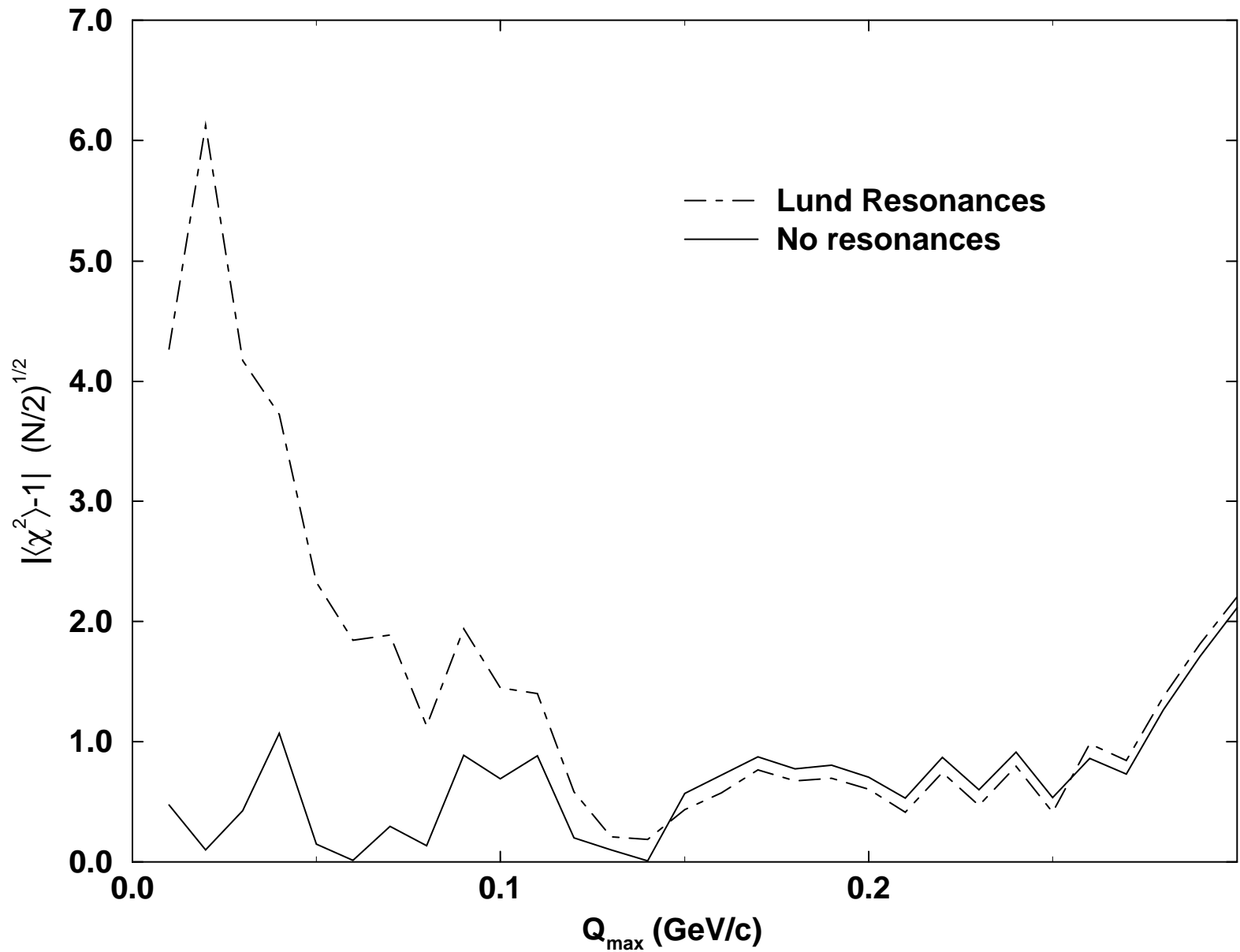
\end{document}